\documentclass[aps, prd, onecolumn, nofootinbib, showpacs]{revtex4}

\usepackage{amsfonts,amsmath,amsthm,amssymb,graphicx}

\usepackage{ulem}
\usepackage[usenames]{color}
\definecolor{DarkGreen}{rgb}{0.0,0.5,0.0}
\usepackage{bm}

\def\spose#1{\hbox to 0pt{#1\hss}}

\def\lta{\mathrel{\spose{\lower 3pt\hbox{$\mathchar"218$}}
     \raise 2.0pt\hbox{$\mathchar"13C$}}}
\def\gta{\mathrel{\spose{\lower 3pt\hbox{$\mathchar"218$}}
     \raise 2.0pt\hbox{$\mathchar"13E$}}}

\begin{document}

\preprint{1234}

\title{Non-Gaussian signatures of Tachyacoustic Cosmology}

\author{Dennis Bessada${}^{1,2}$} \email{dennis.bessada@unifesp.br} \affiliation{
${}^1$UNIFESP - Universidade Federal de S\~ao Paulo -
Laborat\'orio de F\'isica Te\'orica e Computa\c c\~ao
Cient\'ifica,\\ Rua S\~ao Nicolau, 210, 09913-030, Diadema, SP, Brazil\\
              ${}^2$INPE - Instituto Nacional de Pesquisas Espaciais - Divis\~ao de Astrof\'isica, \\Av. dos Astronautas, 1758, 12227-010, S\~ao Jos\'e dos Campos, SP,
              Brazil}

\begin{abstract}

I investigate non-Gaussian signatures in the context of
tachyacoustic cosmology, that is, a noninflationary model with
superluminal speed of sound. I calculate the full non-Gaussian
amplitude $\mathcal{A}$, its size $f_{\rm NL}$, and corresponding
shapes for a red-tilted spectrum of primordial scalar
perturbations. Specifically, for cuscuton-like models I show that
$f_{\rm NL}\sim {\cal O}(1)$, and the shape of its non-Gaussian
amplitude peaks for both equilateral and local configurations, the
latter being dominant. These results, albeit similar, are
quantitatively distinct from the corresponding ones obtained by
Magueijo {\it{et. al}} in the context of superluminal bimetric
models.

\end{abstract}


\maketitle

\section{Introduction}
\label{sec:introduction}

The physics of the very early universe is a rich investigation
field, for there is a plethora of potential models that solve, at
least partially, the well-known problems of the standard
cosmological paradigm. Inflationary cosmology \cite{inflation} is
surely the most successful model, but alternatives to it have been
proposed in the recent past: pre-big bang cosmology
\cite{prebigbang}, ekpyrotic and cyclic models \cite{ekpyrotic},
nonsingular quantum cosmological models \cite{quantumcosmology},
noncanonical models \cite{noncanonical,Bessada:2009ns}, among
others. All of them lead to a near-scale invariant spectrum, so
that a very important question emerges: how can we falsify such
distinct cosmologies? An answer can be provided by the
investigation of their non-Gaussian signatures; single-field
inflation, for example, predicts a nearly-Gaussian CMB anisotropy
field, whereas some noncanonical models, for example, can predict
large deviations from Gaussianity, as in DBI inflation (see, for
example, references \cite{Bessada:2009pe} for large-field
polynomial models, and \cite{Alishahiha:2004eh}). Therefore, the
study of non-Gaussian signatures can be a very important tool to
rule out many models proposed to cope with the problems of the
very early universe.

In particular, noncanonical Lagrangians lead to models with
varying speed of sound, a feature which substantially enhances or
diminishes their non-Gaussian amplitudes, since they are usually
related to terms with a $c_s^{-2}$ dependence
\cite{Seery:2005wm,Chen:2006nt,Khoury:2008wj}; hence, for $c_s<1$,
such terms dominate, enhancing non-Gaussianities as in the case of
DBI inflation. For superluminal models, $c_s\gg 1$, such terms
become subdominant, so that the size of the non-Gaussian amplitude
becomes $f_{\rm NL}\sim 1$ \cite{Magueijo:2010zc,Noller:2011hd},
which distinguishes this class of superluminal models from
inflation (for which $f_{\rm NL}\sim {\cal O}(0.01))$
\cite{Komatsu:2001rj,Maldacena:2002vr}. In \cite{Magueijo:2010zc},
the authors take into account a minimal and a nonminimal bimetric
model, that is, models in which a disformal transformation between
matter and gravity metric is evocated, and has the form
${\hat{g}}_{\mu\nu}=g_{\mu\nu}-B\partial_{\mu}\phi\partial_{\nu}\phi$,
where the coupling $B$ is regarded as a constant in the former,
whereas in the latter it can run with $\phi$. They then calculate
non-Gaussian amplitudes for the nonminimal bimetric model in the
superluminal limit, showing that it only mildly depends on the
tilt $n_s-1$ of the power spectrum. In \cite{Noller:2011hd}, the
authors take a step further and consider non-Gaussianity for a
wider class of models without slow-roll and exact scale invariance
assumptions, including DBI and bimetric models, and study the
superluminal limit of the latter. However, despite the generality
of their analysis, there was still room for another class of
superluminal models, which we presented in \cite{Bessada:2009ns}
and dubbed {\it{tachyacoustic}}.

Tachyacoustic cosmology is a noncanonical, noninflationary model
with superluminal speed of sound, in which a nearly
scale-invariant spectrum is generated by quantum perturbations
redshifted outside of a shrinking acoustic horizon. Since this
model is noninflationary, and the scale factor is a power law of
time, $a(t)\propto t^{1/\epsilon}$, where $\epsilon=const$ is the
flow parameter associated with the ratio $-\dot H/H$, we can take
a radiative equation of state for the model, discarding the need
of a reheating period to make the scalar field decay into
radiation. Also, as shown recently in \cite{Bessada:2012zx}, DBI
and cuscuton-like cosmological solutions also exhibit attractor
behavior. Having this model a nearly scale invariant spectrum of
perturbations, as mentioned, the next step to take is exactly the
analysis of its non-Gaussian features, which is the subject of
this paper.

As we shall see, the distinguishing feature of the tachyacoustic
model with its {\it cuscuton-like} Lagrangian\footnote{A {\it
cuscuton} is a causal field with infinite speed of sound,
originally proposed in \cite{Afshordi:2006ad}. From the flow
hierarchy devised for arbitrary k-essence models
\cite{Bean:2008ga} we reconstructed an action similar to the one
associated with the cuscuton plus a canonical kinetic term, which
accounts for the ``cuscuton-like" term we proposed in
\cite{Bessada:2009ns}.} with regard to the other superluminal
models explored in \cite{Magueijo:2010zc,Noller:2011hd} concerns
the contribution of the term ${\dot{\zeta}^3}$ in the cubic action
\cite{Seery:2005wm,Chen:2006nt} to the full non-Gaussian
amplitude, giving rise to an extra term with a linear dependence
on the flow parameters $\epsilon$ and $s$. This dependence is not
present either in DBI models (since the coefficient of
${\dot{\zeta}^3}$ in the third-order action is identically null in
this case), or in disformal bimetric model, since by projecting
its scalar-field action in the Einstein frame one obtains a
DBI-like action. Therefore, the tachyacoustic model ``inherits" an
extra contribution from the ${\dot{\zeta}^3}$ term, providing a
different value for $f_{\rm NL}$ compared to the other
superluminal models. Despite small, such difference might be
observable in the near future, what is of great importance for
falsifying superluminal noncanonical models.

This paper is organized as follows. In Section \ref{sec:basics} I
review the basics of tachyacoustic cosmology, and in Section
\ref{sec:perturbations} I derive some important results concerning
the power spectrum. In Section \ref{sec:3orderaction} I review the
third-order action in the fluctuations, whereas in
\ref{sec:ngamplitudes} I calculate the $\dot\zeta^3$ contribution
to the non-Gaussian amplitudes, which is absent in other models.
In Sections \ref{sec:sizengamplitudes} and \ref{sec:shape} I
discuss, respectively, the size and the shapes of the non-Gaussian
amplitudes for the cuscuton-like tachyacoustic model. In
\ref{sec:concl} I summarize the main results of this paper.

\section{Tachyacoustic Cosmology}
\label{sec:tachyac}

\subsection{Basics}
\label{sec:basics}

One of the crucial problems of the very-early cosmology is the
presence of a horizon that prevents a causal explanation for the
observed features of the universe. The inflationary paradigm
tackles this problem by means of a huge accelerated expansion in
the very early universe, enforcing physical scales to be deep
inside the horizon and, in consequence, in causal contact.
However, if light could have travelled much faster in the early
universe, all scales would have been naturally in causal contact,
so that the horizon problem would not be an issue at all. This is
the basic proposal of the varying speed of light theories (VSL)
\cite{vsl}. Despite VSL theories fail to explain cosmic
structures, their essential idea remains a very insightful
approach to search for alternatives to the inflationary paradigm.

Along with the speed of light, another velocity parameter comes
into play when studying cosmological perturbations: the speed of
sound parameter $c_s$. In the standard inflationary scenario,
$c_s$ is a constant and coincides with the speed of light.
However, in k-essence models, the presence of a noncanonical
kinetic term gives rise to the possibility of a varying speed of
sound. Also, such models possess two horizons, a Hubble horizon
and an acoustic horizon, which have independent dynamics; since
the acoustic horizon depend on the speed of sound, it can be large
or even infinite at early times for large values of $c_s$, solving
the cosmological horizon problem without inflation. In this case,
as in VSL models, superluminal propagation is again evoked, for
$c_s$ represents the speed of propagation of the perturbations on
a nontrivial background; then, a relevant question now arises: are
these models causally consistent?

Fortunately, causal problems are usually model-dependent. As
pointed out in references \cite{Bruneton:2006gf} and
\cite{Babichev:2007dw}, causal problems do not affect general
k-essence models or bimetric theories; also, cuscuton models are
free from them \cite{Afshordi:2006ad}. In particular, it can be
shown that the noncanonical structure of a general k-essence model
induces an ``acoustic" metric ${{G}}^{\mu\nu}$
\cite{Babichev:2007dw}; hence, this two-metric model possess two
distinct local causal structures, the first one being the usual
light cone (connected with the background metric $g_{\mu\nu}$ of
the manifold ${\cal{M}}$), and the second one connected to the
speed of sound (thereby the ``acoustic cone"). It can be shown
that if there exists a time coordinate $t$ with respect to the
background metric (which is everywhere future directed) which
plays a role of global time for {\it{both}} spacetimes
$\left({\cal{M}},g_{\mu\nu}\right)$ and
$\left({\cal{M}},G^{-1}_{\mu\nu}\right)$, which guarantees the
absence of closed causal curves  for k-essence models built from
homogeneous scalar fields on a FRW background
\cite{Babichev:2007dw}. Since the causal structure for small
perturbations is determnined by the acoustic cone, we can study
superluminal propagation safely, for no causal paradoxes arise.

Tachyacoustic cosmology is a particular solution of a wider class
of k-essence models (see Bean {\it{et al.}} \cite{Bean:2008ga}),
which we briefly review below. Consider a general Lagrangian of
the form ${\cal{L}}={\cal{L}}\left[X,\phi\right]$, where
$2X=g^{\mu\nu}\partial_{\mu}\phi\partial_{\nu}\phi$ is the
canonical kinetic term ($X>0$ according to our choice of the
metric signature). The energy density $\rho$ and pressure $p$ are
given by
\begin{eqnarray}
P &=& {\cal L}\left(X,\phi\right),\label{defP}\\
\rho &=& 2 X {\cal L}_X - {\cal L}\label{defrho},
\end{eqnarray}
whereas the speed of sound is given by
\begin{eqnarray}
\label{defspeedofsound} c_s^{2} &\equiv& \frac{P_X}{\rho_X} =
\left(1 + 2X\frac{{\cal L}_{XX}}{{\cal L}_{X}}\right)^{-1},
\end{eqnarray}
where the subscript ``${X}$" indicates a derivative with respect
to the kinetic term. The Friedmann equation can be written in
terms of the reduced Planck mass  $M_P=1/\sqrt{8\pi G}$
\begin{equation}
\label{eqFriedmann} H^2 = \frac{1}{3 M_P^2} \rho = \frac{1}{3
M_P^2} \left(2X {\cal L}_X - {\cal L}\right).
\end{equation}

For monotonic field evolution, the field value $\phi$ can be used
as a ``clock'', and all other quantities expressed as functions of
$\phi$, for example $X = X\left(\phi\right)$, ${\cal L} = {\cal
L}\left[X\left(\phi\right),\phi\right]$, and so on. We consider
the homogeneous case, so that $\dot\phi = \sqrt{2 X}$. Next, using
\begin{equation}
\frac{d}{dt} = \dot\phi \frac{d}{d \phi} = \sqrt{2 X}
\frac{d}{d\phi},
\end{equation}
we can re-write the Friedmann equation as the {\it Hamilton
Jacobi} equation
\begin{eqnarray}
\label{hamjac1} \dot \phi = \sqrt{2 X} &=& -\frac{2M_P^2}{{\cal
L}_{X}}H'(\phi),
\end{eqnarray}
where a prime denotes a derivative with respect to the field
$\phi$. The number of e-folds $dN$ can similarly be re-written in
terms of $d\phi$ by:
\begin{eqnarray}
\label{eq:Nphi} dN \equiv -H dt = - \frac{H}{\sqrt{2 X}} d\phi=
\frac{{\cal L}_X}{2 M_P^2}
\left(\frac{H\left(\phi\right)}{H'\left(\phi\right)}\right) d\phi.
\end{eqnarray}

We introduce a hierarchy of flow parameters, the first three
defined by
\begin{eqnarray}
\label{defeps} \epsilon\left(\phi\right) &\equiv& -\frac{\dot
H}{H^2}= \frac{2 M_P^2}{{\cal{L}}_{X}}
\left(\frac{H'\left(\phi\right)}{H\left(\phi\right)}\right)^2,\\
\label{defs} s\left(\phi\right) &\equiv& \frac{{\dot{c}}_s}{c_sH}=
- \frac{2 M_P^2}{{\cal{L}}_{X}}
\frac{H'\left(\phi\right)}{H\left(\phi\right)}
\frac{c_S'\left(\phi\right)}{c_S\left(\phi\right)},\\
\label{defstil} \tilde{s}\left(\phi\right) &\equiv&
-\frac{{\dot{\mathcal{L}}}_X}{H{\mathcal{L}}_X} = \frac{2
M_P^2}{{\cal{L}}_{X}}
\frac{H'\left(\phi\right)}{H\left(\phi\right)}\frac{{\cal{L'}}_{X}}{{\cal{L}}_{X}},
\end{eqnarray}
where we have used equation (\ref{hamjac1}). Note that using
definitions (\ref{eqFriedmann}), (\ref{defeps}), we can write down
the equation of state for a k-essence fluid as
\begin{eqnarray}
\label{kessstateq} P&=&\left(\frac{2}{3}\epsilon-1\right)\rho,
\end{eqnarray}
where we have used equations (\ref{defP}) and (\ref{defrho}).
Therefore, the equation of state parameter $w$ can be written in
terms of $\epsilon$ and, conversely,
\begin{eqnarray}
\epsilon&=&\frac{3}{2}\left(1+w\right).
\end{eqnarray}
Hence, a matter-dominated ($w=0$) tachyacoustic model has
$\epsilon=3/2$, whereas a radiation dominated one ($w=1/3$) has
$\epsilon=2$.

In the case of Tachyacoustic Cosmology, we take all flow
parameters to be constant, and then solve the flow hierarchy,
reconstructing the action at the end. In particular, we find that
the Hubble parameter and the speed of sound evolve as
\begin{equation}
\label{Hubble}
H\left(\phi\right)=H_0\left(\frac{\phi}{\phi_0}\right)^{\epsilon/s},
\end{equation}
\begin{equation}
\label{c_s} c_S\left(\phi\right)=\frac{\phi_0}{\phi},
\end{equation}
and taking $\tilde{s}=-2s$, the resulting Lagrangian has a
cuscuton-like form \cite{Bessada:2009ns}
\begin{equation}
\label{lagcuscs}
{\cal{L}}\left(X,\phi\right)=2f\left(\phi\right)\sqrt{X} + C X -
V\left(\phi\right),
\end{equation}
where $f\left(\phi\right)$ is the analog of the inverse brane
tension, $V\left(\phi\right)$ is the potential and $C$ is a
constant given by
\begin{equation}
\label{defC} C=\frac{2M_P^2\epsilon}{s^2\phi_0^2}.
\end{equation}
To conclude this review, is important to mention that, as pointed
out at the beginning of this section and demonstrated in
\cite{Bessada:2009ns}, tachyacoustic cosmology is free from causal
paradoxes. Also, as the acoustic horizon is given by
\begin{equation}
D_H \simeq \frac{c_S}{a H},
\end{equation}
we can be shown that its conformal-time dependency goes like
\begin{equation}
D_H \propto \frac{c_S}{a H} \propto e^{\left(1 - \epsilon -
s\right) N} \propto \tau^{\left(1 - \epsilon - s\right) / \left(1
- \epsilon\right)}.
\end{equation}
Therefore the condition for a shrinking acoustic horizon, $1 -
\epsilon - s > 0$, is {\it not} identical to accelerated
expansion. For $\epsilon > 1$ and $s < 1 - \epsilon$, the
expansion is non-inflationary, the Hubble horizon is growing in
comoving units, and the acoustic horizon is shrinking. The initial
singularity is at $\tau = 0$, and we see immediately that for the
tachyacoustic solution, the speed of sound in the scalar field is
{\it infinite} at the initial singularity, and the acoustic
horizon is likewise infinite in size. Therefore, such a cosmology
presents no ``horizon problem'' in the usual sense, since even a
spatially infinite spacetime is causally connected on the
initial-time boundary.

There are two other important parameters to be taken into account
in our discussion. They appear as the coefficient of the term
${\dot{\zeta}^3}$ when one expands a general noncanonical action
up to third-order in the fluctuations, in the following
combination:
\begin{eqnarray}
\label{defLambda}
\Lambda\equiv\Sigma\left(1-\frac{1}{c_s^2}\right)+2\lambda,
\end{eqnarray}
where the parameters $\Sigma$ and $\lambda$ are given by
\cite{Seery:2005wm,Chen:2006nt},
\begin{eqnarray}
\label{defSigma}
\Sigma&=&X {\cal{L}}_{X}+2X^2{\cal{L}}_{XX}  = \frac{H^2\epsilon}{c_s^2},\\
\label{deflambda} \lambda&=&
X^2{\cal{L}}_{XX}+\frac{2}{3}X^3{\cal{L}}_{XXX}.
\end{eqnarray}
The parameter $\lambda$ can also be written as
\cite{Seery:2005wm,Khoury:2008wj}
\begin{eqnarray}
\label{lambdaf} \lambda = \frac{\Sigma}{6}\left( \frac{2 f_X +
1}{c_s^2} -1 \right),
\end{eqnarray}
where $f_X$ and the ``kinetic part" $\epsilon_X$ of the flow
parameter $\epsilon=\epsilon_X+\epsilon_{\phi}$, are respectively
given by
\begin{eqnarray}
f_X = \frac{\epsilon s}{3 \epsilon_X}, \qquad \epsilon_X =
-\frac{\dot{X}}{H^2}\frac{\partial H}{\partial X}.
\end{eqnarray}

The $\Sigma$ term as presented in the right-hand side of equation
(\ref{defSigma}) holds the same form regardless of the
noncanonical model involved, but the $\lambda$ term may vary
according to the underlying Lagrangian. For DBI-like Lagrangians,
for instance, $\lambda$ is given exactly by \cite{Chen:2006nt}
\begin{eqnarray}
\label{lambdaDBI} \lambda_{DBI}=
\frac{H^2\epsilon}{2c_s^4}\left(1-c_s^2\right),
\end{eqnarray}
whereas for an arbitrary noncanonical model expression
(\ref{lambdaf}) is generally employed with the assumption that
$f_X$ is constant \cite{Khoury:2008wj}. Again, for DBI-like
models, from (\ref{defLambda}), (\ref{defSigma}) and
(\ref{lambdaDBI}) it is straightforward to show that
\begin{eqnarray}
\label{LambdaDBI} \Lambda_{DBI}=0,
\end{eqnarray}
so that the ${\dot{\zeta}^3}$ contribution to the action vanishes
identically. This fact accounts for the absence of the
${\mathcal{A}}_{\dot{\zeta}^3}$ contribution to the three-point
function amplitude of DBI models. The same happens to disformal
bimetric models, as argued in the Introduction, for their action
reduce to a DBI-like one when projected in the Einstein frame.

In the cuscuton-like Lagrangian (\ref{lagcuscs}), we do not have
to worry about $\lambda$, for it is identically null:
\begin{eqnarray}
\label{lambdaTac} \lambda_{cusc}=0,
\end{eqnarray}
so that
\begin{eqnarray}
\label{Lambdatach}
\Lambda_{cusc}=\frac{H^2\epsilon}{c_s^2}\left(1-\frac{1}{c_s^2}\right);
\end{eqnarray}
hence, the coefficient of the ${\dot{\zeta}^3}$-term does not
vanish, unlike the superluminal models described above, yielding a
${\cal A}_{\dot\zeta^3}$ contribution to the amplitude.

\subsection{Perturbations}
\label{sec:perturbations}

In \cite{Bessada:2009ns} we have solved the mode equation for an
arbitrary noncanonical model \cite{Garriga:1999vw} in the case of
constant flow parameters (as defined in equations
(\ref{defeps}-\ref{defstil})). In this paper I will take a
slightly different path, using the so-called ``sound-horizon" time
${\rm d}y = c_s {\rm d}\tau$ instead of the conformal time $\tau$
\cite{Khoury:2008wj}, just for the sake of comparisons with the
results obtained in \cite{Noller:2011hd}.

Taking $\epsilon$, $s$, $\tilde{s}$ constant, using definitions
(\ref{defeps},\ref{defs}), and defining
\begin{eqnarray}
\label{defgam} \gamma \equiv \epsilon+s -1,
\end{eqnarray}
it follows that
\begin{eqnarray}
\label{soundhortime} y = \frac{c_s}{\gamma aH},
\end{eqnarray}
and
\begin{eqnarray} \label{acsHy}
a \sim (-y)^{1/\gamma}, \;\; c_s \sim (-y)^{s/\gamma}, \;\; H \sim
(-y)^{-\epsilon/\gamma}.
\end{eqnarray}

We next turn to the second-order action for the curvature
perturbation $\zeta$ \cite{Garriga:1999vw}:
\begin{eqnarray}
\label{2ordaction} S_2 = \frac{M_{P}^2}{2} \int {\rm d}^3x{\rm
d}\tau \;z^2\left[ \left(\frac{d\zeta}{d\tau}\right)^2 -
c_s^2({\nabla}\zeta)^2\right],
\end{eqnarray}
where $z$ is defined as $z = a \sqrt{2 \epsilon}/c_s$. From now
on, a prime $'$ indicates a derivative with respect to $y$; then,
defining
\begin{eqnarray}
\label{defq}
q \equiv \sqrt{c_s}z,
\end{eqnarray}
action (\ref{2ordaction}) becomes
\begin{eqnarray}
\label{yact}
S_2 = \frac{M_{P}^2}{2}\int {\rm d}^3x{\rm d}y
\;q^2\left[ \zeta'^2 -({\nabla}\zeta)^2\right];
\end{eqnarray}
next, introducing the canonical variable $v=M_P q\zeta$, from
(\ref{yact}) we derive the mode equation
\begin{eqnarray}
\label{modeq}
v''_k +\left(k^2 - \frac{q''}{q}\right)v_k = 0.
\end{eqnarray}

After a little algebra we can rewrite the last term in the
equation above as
\begin{equation}
\label{derq}
\frac{q''}{q} = \frac{1}{y^2}\left(\nu^2 -
\frac{1}{4}\right),
\end{equation}
where
\begin{equation}
\label{defnu}
\nu\equiv \frac{\epsilon+2s-3}{2(\epsilon+s-1)}.
\end{equation}
The scale-invariant limit is achieved when $q''/q = 2/y^2$, so
that $\nu=3/2$ in this case. Since curvature perturbations $\zeta$
are quantum fields, we expand them in terms of creation and
annihilation operators,
\begin{equation}
\label{qmodes}
\zeta(y, \mathbf{k}) = u_k(y)a_{\mathbf{k}} +
u_k^*(y) a^\dagger_{-\mathbf{k}},
\end{equation}
so that the solution for $v_k(y)$ associated with the Bunch Davis
vacuum is
\begin{equation}
\label{vkmodes}
v_k(y) = \frac{\sqrt{\pi}}{2} \sqrt{- y} \,
H^{(1)}_\nu (- k y),
\end{equation}
where $H^{(1)}_\nu$ are Hankel functions of the first kind. Next,
from the modes introduced in (\ref{qmodes}) and solution
(\ref{vkmodes}) we find that
\begin{eqnarray}
\label{expuk} u_k(y) = \frac{c_s^{1/2}}{a M_P 2^{3/2}}
\sqrt{\frac{\pi}{\epsilon}} \sqrt{-y} H^{(1)}_\nu (- k y) \approx
-  \, \frac{iH \gamma}{2 M_P \sqrt{ c_s k^3 \epsilon}}
\left(\frac{- k y}{2}\right)^{3/2 - \nu} (1 + i k y) e^{- i k y},
\end{eqnarray}
where we have used the following expansion of Hankel functions for
small arguments, $|k y| \ll 1$,
\begin{equation}
\label{Hankel}
H^{(1)}_\nu (- k y) = - i \, \frac{2^\nu
\Gamma(\nu) (- k y)^{-\nu}}{\pi} [1+ i k y + {\cal O}(k y)^2]
e^{-i k y},
\end{equation}
together with the assumption that $\Gamma(\nu \approx 3/2) \approx
\sqrt{\pi}/2$, which corresponds to nearly scale invariance, as
desired. From expression (\ref{Hankel}) we can also compute the
derivative of $u_k(y)$ with respect to $y$,
\begin{equation}
\label{deruk} u'_k(y) \approx -i \, \frac{H (\epsilon + s -1)}{2
M_P \sqrt{ c_s k^3 \epsilon}} \left(\frac{- k y}{2}\right)^{3/2 -
\nu} k^2 y \ e^{- i k y},
\end{equation}
which is plays an important role in the computations of section
\ref{sec:ngamplitudes}.

From the definition of the $\zeta$ power spectrum
\begin{equation}
P_\zeta \equiv \frac{1}{2\pi^2}k^3\left\vert\zeta_k\right\vert^2,
\end{equation}
we find,
\begin{equation}
\label{Pz} P_\zeta = \frac{(\epsilon + s -1)^2 \, 2^{2 \nu - 3}}{2
(2 \pi)^2M_{P}^2\epsilon}\frac{{\bar H}^2}{{\bar c}_s},
\end{equation}
where the bar refers to quantities evaluated at sound
horizon-crossing, $y = k^{-1}$. From the computation of the power
spectrum we find the expression for the scalar spectral index
$n_s$, which is related to $\nu$ and the flow parameters
$\epsilon$ and $s$ by
\begin{equation}
\label{nsexp} n_s - 1 = 3 - 2 \nu = \frac{2 \epsilon + s}{s +
\epsilon -1}.
\end{equation}

\section{Non-Gaussian Signatures}
\label{sec:nongaussign}

\subsection{Third-order action}
\label{sec:3orderaction}

In order to investigate non-Gaussian signatures, we basically
compute the three-point function for the curvature perturbations,
starting from the third-order action in the ADM formalism
\cite{Seery:2005wm,Chen:2006nt}:
\begin{eqnarray}
\label{3ordaction} S_3&=&{M_P}^2\int {\rm d}t \, {\rm d}^3x
\left\{ -a^3
\left[\Sigma\left(1-\frac{1}{c_s^2}\right)+2\lambda\right]
\frac{\dot{\zeta}^3}{H^3}
+\frac{a^3\epsilon}{c_s^4}(\epsilon-3+3c_s^2)\zeta\dot{\zeta}^2-
2a \frac{\epsilon}{c_s^2}\dot{\zeta}(\partial \zeta)(\partial
\chi) \right. \nonumber \\ &+&
\frac{a\epsilon}{c_s^2}(\epsilon-2s+1-c_s^2)\zeta(\partial\zeta)^2+\frac{a^3\epsilon}{2c_s^2}\frac{d}{dt}\left(\frac{\eta}{c_s^2}\right)\zeta^2\dot{\zeta}
+\frac{\epsilon}{2a}(\partial\zeta)(\partial \chi) \partial^2 \chi
+ \frac{\epsilon}{4a}(\partial^2\zeta)(\partial \chi)^2\nonumber
\\ &+& \left. 2 f(\zeta)\left.\frac{\delta L}{\delta
\zeta}\right\vert_1 \right\};
\end{eqnarray}
such expression is valid outside of the slow-roll approximation
and for any time-independent sound speed. In (\ref{3ordaction})
the dots denote derivatives with respect to proper time $t$, and
$\chi$ is defined as
\begin{equation}
\partial^2 \chi = \frac{a^2 \epsilon}{c_s^2}\dot{\zeta},
\end{equation}
whereas the elements appearing in the last term of the action
(\ref{3ordaction}) are
\begin{eqnarray}
\label{fzeta}
f(\zeta)&=&\frac{\eta}{4c_s^2}\zeta^2+\frac{1}{c_s^2H}\zeta\dot{\zeta}+
\frac{1}{4a^2H^2}[-(\partial\zeta)(\partial\zeta)+\partial^{-2}(\partial_i\partial_j(\partial_i\zeta\partial_j\zeta))]+\frac{1}{2a^2H}[(\partial\zeta)(\partial\chi)-\partial^{-2}(\partial_i\partial_j(\partial_i\zeta\partial_j\chi))],
%\nonumber \\
%&+&
\end{eqnarray}
\begin{eqnarray}
\label{dLdzeta}
\left.\frac{\delta L}{\delta\zeta}\right\vert_1
&=& a \left( \frac{d\partial^2\chi}{dt}+H\partial^2\chi
-\epsilon\partial^2\zeta \right),
\end{eqnarray}
where $\partial^{-2}$ is the inverse Laplacian. Note that
(\ref{dLdzeta}) corresponds to the linearized equations of motion,
and can be absorbed by a field redefinition
\begin{eqnarray}
\zeta \rightarrow \zeta_n+f(\zeta_n).
\end{eqnarray}

Since in the limit $y\rightarrow 0$ one has
\begin{equation}
\frac{H}{c_s^{1/2}} \sim (-y)^{\nu - 3/2},
\end{equation}
it is clear from expression (\ref{expuk}) that
$\zeta\longrightarrow const$ on superhorizon scales, so that we
can drop the second term in (\ref{fzeta}), and we are left with
the field redefinition $\zeta \rightarrow
\zeta_n+\eta\zeta_n^2/(4c_s^2)$; in the present case, for constant
flow parameters we get $\eta=0$, so that we can simply skip the
last term in the action (\ref{3ordaction}).

\subsection{Amplitudes}
\label{sec:ngamplitudes}

I next turn to the calculation of the three-point function which
leads to the $f_{\rm NL}$ parameter, following the same steps as
taken in \cite{Khoury:2008wj}. In the interaction picture, the
three-point function is given by
\begin{equation}
\label{defthreepoint} \langle
\zeta(t,\textbf{k}_1)\zeta(t,\textbf{k}_2)\zeta(t,\textbf{k}_3)\rangle=
-i\int_{t_0}^{t}{\rm d}t^{\prime}\langle[
\zeta(t,\textbf{k}_1)\zeta(t,\textbf{k}_2)\zeta(t,\textbf{k}_3),H_{\rm
int}(t^{\prime})]\rangle,
\end{equation}
where $H_{\rm int}=-L_{\rm int}$ is the interaction Hamiltonian
derived from the corresponding Lagrangian (\ref{3ordaction}), and
vacuum expectation values are evaluated with respect to the
interacting vacuum $|\Omega \rangle$ (a through discussion on this
point can be found in \cite{Seery:2005wm}); $t_0$ is some
sufficiently early time. Except from the $\dot\zeta^3$ term, as
discussed in the Introduction, all the amplitudes derived have the
same form as the amplitudes evaluated in \cite{Noller:2011hd}.
Then, I shall evaluate here only the ${\cal A}_{\dot\zeta^3}$
contribution to the non-Gaussian amplitude, which is derived from
\begin{eqnarray}
\label{dotzeta3}
S_{\dot{\zeta}^3}&=&{M_P}^2\int^{y_{\rm
end}}_{-\infty+i\varepsilon} {\rm d}y \, {\rm d}^3x
\frac{c_s^3}{H^3} \Sigma\left(1-\frac{1}{c_s^2}\right)
{{\zeta\,'}^3},
\end{eqnarray}
where we have changed the time variable $t$ with the sound-horizon
time variable $dy=c_s dt/a$, and used (\ref{lambdaTac}). The
parameter $y_{\rm end}<0$ indicates the ``end" of the deceleration
period with infinite speed of sound (in \cite{Khoury:2008wj}, it
refers to the end of inflationary or ekpyrotic phase). Next,
substituting contribution (\ref{qmodes}) and (\ref{dotzeta3}) into
(\ref{defthreepoint}), and using the commutation relations
$[a(\mathbf{k}), a^\dagger(\mathbf{k}')] = (2 \pi)^3
\delta^{(3)}(\mathbf{k} - \mathbf{k}')$, we find
\begin{multline}
\label{z33point1} \langle
\zeta(\textbf{k}_1)\zeta(\textbf{k}_2)\zeta(\textbf{k}_3)\rangle_{
\dot\zeta^3}\, =\, - (2 \pi)^3i
\delta^{(3)}(\mathbf{k}_1+\mathbf{k}_2+\mathbf{k}_3)
u_{k_1}(y_{\rm end})u_{k_2}(y_{\rm end})u_{k_3}(y_{\rm end})
\\\times\left\{  \int_{-\infty+i\varepsilon}^{y_{\rm end}} {\rm d}
y \frac{ac_s^2}{H^3}\Sigma\left(1-\frac{1}{c_s^2}\right)
u_{k_1}^*(y)' u_{k_2}^*(y)' u_{k_3}^*(y)' + {\rm perm.} + {\rm
c.c.}\right\}.
\end{multline}
Also, substituting (\ref{defSigma}), (\ref{soundhortime}),
(\ref{expuk}) and (\ref{deruk}) into equation (\ref{z33point1}),
we find
\begin{multline}
\label{z33point2}
\langle
\zeta(\textbf{k}_1)\zeta(\textbf{k}_2)\zeta(\textbf{k}_3)\rangle_{
\dot\zeta^3}\, =\, - (2 \pi)^3i
\delta^{(3)}(\mathbf{k}_1+\mathbf{k}_2+\mathbf{k}_3)
\frac{\gamma^6}{2^{15-6\nu}M_P^4\epsilon^3}\frac{H_{\rm
end}^3}{c_{s,{\rm
end}}^{3/2}}\left(k_1k_2k_3\right)^{2-2\nu}\left\vert y_{\rm
end}\right\vert^{9/2-3\nu}
\\ \times
\frac{\epsilon}{\gamma}\int_{-\infty+i\varepsilon}^{y_{\rm end}}
{\rm d} y \frac{H}{c_{s}^{1/2}}
\left(1-\frac{1}{c_s^2}\right)y^2y^{9/2-3\nu}e^{iKy}
+ {\rm perm.} + {\rm c.c.}\,,
\end{multline}
where $K\equiv k_1+k_2+k_3$. The integral appearing in the
expression can be simplified as follows: first, from (\ref{acsHy})
we get the following expressions,
\begin{equation}
\label{Hcs}
\frac{H}{c_s^{1/2}} = \frac{H_{\rm end}}{c_{s,\rm
end}^{1/2}}\left(\frac{y}{y_{\rm end}}\right)^{\alpha},\qquad
\frac{H}{c_s^{5/2}}\left(\frac{y}{y_{\rm end}}\right)^{\nu - 3/2}
= \frac{H_{\rm end}}{c_{s,\rm end}^{5/2}}\left(\frac{y}{y_{\rm
end}}\right)^{\beta},
\end{equation}
where we have defined the exponents
\begin{eqnarray}
\label{defexpon}
\alpha \equiv 3 - 2\nu = \frac{2 \epsilon +
s}{\epsilon + s - 1}, \qquad \qquad \beta  \equiv  \frac{2
\epsilon - s}{\epsilon + s -1};
\end{eqnarray}
next, defining
\begin{equation}
\label{defint}
{\cal I}_n(p) \equiv {\rm Im}\left\{\int_{-\infty +
i\varepsilon}^{y_{\rm end}} {\rm d} y \left(\frac{y}{y_{\rm
end}}\right)^p (- i y)^n e^{i K y}\right\},
\end{equation}
we can reduce expression (\ref{z33point2}) to
\begin{multline}
\label{z33point3} \langle
\zeta(\textbf{k}_1)\zeta(\textbf{k}_2)\zeta(\textbf{k}_3)\rangle_{
\dot\zeta^3}\, =\,  (2 \pi)^3i
\delta^{(3)}(\mathbf{k}_1+\mathbf{k}_2+\mathbf{k}_3)
\left(k_1k_2k_3\right)^{2-2\nu}\frac{\gamma^5}{2^{15-6\nu}M_P^4\epsilon^2}\frac{H_{\rm
end}^4}{c_{s,{\rm end}}^{2}}\left\vert y_{\rm
end}\right\vert^{9-6\nu}
\\ \times
\left[\frac{1}{c_{s,{\rm end}}^{2}}{\cal I}_2(\beta)-{\cal
I}_2(\alpha)\right]
+ {\rm perm.} + {\rm c.c.}\,.
\end{multline}

It is important to stress that we took the imaginary part of
integral (\ref{defint}) for this is the only part relevant for
this calculation \cite{Khoury:2008wj}. For all modes of interest,
$K |y_{\rm end}|$ is a small quantity; then, as discussed
in~\cite{Maldacena:2002vr}, the above choice of integration
contour picks up the appropriate interacting ``in-in" vacuum at
$|y|\rightarrow \infty$ and takes care of the oscillating behavior
of the exponential.

We can further simplify expression (\ref{z33point3}) by relating
the quantities evaluated at $y=y_{\rm end}$ to their corresponding
values at sound horizon-crossing, $y=K^{-1}$,
\begin{equation}
\label{Hcsendbar}
H_{\rm end} = \bar H\left(Ky_{\rm
end}\right)^{-\epsilon/\gamma},\qquad
c_{s,\rm end} = {\bar c}_s\left(Ky_{\rm end}\right)^{s/\gamma},
\end{equation}
so that we use the expression for the $\zeta$ power-spectrum
(\ref{Pz}) to absorb the constants $\bar H$ and ${\bar c}_s$ in
(\ref{z33point3}); the final result is
\begin{equation}
\label{defz3amplit}
\langle
\zeta(\textbf{k}_1)\zeta(\textbf{k}_2)\zeta(\textbf{k}_3)\rangle_{
\dot\zeta^3} = (2\pi)^7 \delta^3({\mathbf k}_1+{\mathbf
k}_2+{\mathbf k}_3) P_\zeta^{\;2} \frac{1}{\Pi_j k_j^3}{\cal A}_{
\dot\zeta^3},
\end{equation}
where
\begin{equation}
\label{z33point}
{\cal A}_{\dot\zeta^3}\, = -
\frac{3}{4}\frac{\epsilon + s - 1}{ {\bar c}_s^{2}}
\left(\frac{k_1k_2k_3}{2 K^3}\right)^{n_s - 1}
\left[I_{\dot\zeta^3}(\beta) - {\bar c}_s^2
I_{\dot\zeta^3}(\alpha) \right],
\end{equation}
and
\begin{equation}
\label{defintz3}
I_{\dot\zeta^3}(p)\equiv -
\left(k_1k_2k_3\right)^2\left(K\left\vert y_{\rm
end}\right\vert\right)^p {\cal I}_2(p).
\end{equation}

Using the same methods the other relevant integrals are given by
(they coincide with the ones computed in \cite{Noller:2011hd}):
\begin{eqnarray}
\label{otamplit1}
{\cal A}_{\zeta \dot\zeta^2} &=& \frac{1}{4 {\bar
c}_s^{2}} \left(\frac{k_1k_2k_3}{2 K^3}\right)^{n_s - 1}
\left[(\epsilon - 3) I_{\zeta \dot\zeta^2}(\beta) + 3 {\bar c}_s^2
I_{\zeta \dot\zeta^2}(\alpha)   \right], \\
\label{otamplit2}
{\cal A}_{\zeta(\partial \zeta)^2} &=& \frac{1}{8
{\bar c}_s^{2}} \left(\frac{k_1k_2k_3}{2 K^3}\right)^{n_s - 1}
\left[(\epsilon - 2 s +1) I_{\zeta(\partial \zeta)^2}(\beta) -
{\bar c}_s^2 I_{\zeta(\partial \zeta)^2}(\alpha)   \right],
\end{eqnarray}
where
\begin{eqnarray}
\label{defotint1}
{I}_{\zeta \dot\zeta^2}(p) &\equiv&
-\left(K\left\vert y_{\rm
end}\right\vert\right)^p\left\{\left[{\cal I}_0(p)+K{\cal
I}_1(p)\right]\sum_{i<j}k_i^2k_j^2 -  {\cal I}_1(p)\frac{1}{K^2}
\sum_{i\neq j} k_i^2 k_j^3\right\},
\\
\label{defotint2}
{I}_{\zeta(\partial \zeta)^2}(\alpha) &\equiv&
\left(K\left\vert y_{\rm end}\right\vert\right)^p
\left(\sum_i k_i^2\right)\left[{\cal I}_{-2}(p)+K{\cal I}_{-1}(p)
+ {\cal I}_{0}(p) \sum_{i<j}k_i k_j + k_1 k_2 k_3{\cal
I}_{1}(p)\right].
\end{eqnarray}

Contributions ${\cal A}_{\dot \zeta \partial \zeta \partial
\chi}$, ${\cal A}_{\partial \zeta
\partial \chi \partial^2 \chi}$ and ${\cal A}_{(\partial^2
\zeta) (\partial \chi)^2}$ to the full amplitude, as found in
\cite{Noller:2011hd}, are subdominant, for they go to zero as
$\bar{c}_s\rightarrow\infty$, so that we neglect them here.

Integrals of the form (\ref{defint}) are convergent for $p+n>-2$
as $y_{\rm end}\rightarrow 0$ \cite{Khoury:2008wj}, and assume the
form
\begin{equation} \label{Cconvergent}
{\cal I}_n(p) = - (K |y_{\rm end}|)^{-p} \cos\frac{p \pi}{2}
\Gamma(1+ p + n) K^{-n-1};
\end{equation}
then, in the limit $\bar{c}_s \to \infty$, the relevant integrals
(\ref{defintz3}), (\ref{defotint1}) and (\ref{defotint2}) assume
the form
\begin{eqnarray}
\label{intcsinf1} I_{\dot\zeta^3}(\alpha)&=&
\frac{\left(k_1k_2k_3\right)^2}{K^3}\cos\frac{\alpha \pi}{2}
\Gamma(3 + \alpha), \\
\label{intcsinf2} {I}_{\zeta \dot\zeta^2}(\alpha) &=&
\cos\frac{\alpha \pi}{2} \Gamma(1 + \alpha)
\left[ \frac{(2 + \alpha)}{K}\sum_{i<j}k_i^2k_j^2 - \frac{(1 +
\alpha)}{K^2}\sum_{i\neq j} k_i^2 k_j^3\right],
\\
\label{intcsinf3}
{I}_{\zeta(\partial \zeta)^2}(\alpha) &=&
-\cos\frac{\alpha \pi}{2}\Gamma(1 + \alpha)\left(\sum_i
k_i^2\right)\left[\frac{K}{\alpha - 1} + \frac{1}{K}\sum_{i<j}k_i
k_j + (1 + \alpha)\frac{k_1 k_2 k_3}{K}\right],
\end{eqnarray}
where $\alpha$ is given by (\ref{defexpon}). Adding up all the
relevant contributions to the full amplitude ${\cal A} = {\cal
A}_{\dot\zeta^3} + {\cal A}_{\zeta \dot\zeta^2} + {\cal
A}_{\zeta(\partial \zeta)^2}$, given by (\ref{z33point}),
(\ref{otamplit1}) and (\ref{otamplit2}) respectively, taking the
small tilt ($n_s - 1 \ll 1$) and $\bar{c}_s \to \infty$ limits, we
find that the full amplitude for cuscuton-like models is given by
\begin{align}
\label{ngamplitude} {\cal A}_{\bar{c}_s \to \infty} \ = &\
\left(\frac{k_1 k_2 k_3}{2 K^3}\right)^{n_s -1}
\left\{\frac{3}{2}\left(\epsilon + s -
1\right)\frac{\left(k_1k_2k_3\right)^2}{K^3}
- \frac{1}{8}\sum_i k_i^3 + \frac{1}{K} \sum_{i<j} k_i^2 k_j^2 -
\frac{1}{2 K^2} \sum_{i\neq j}k_i^2 k_j^3\, \right. \nonumber \\
&\ + (n_s -1)\left.\left[
\frac{9}{4}\left(\epsilon + s -
1\right)\frac{\left(k_1k_2k_3\right)^2}{K^3}
- \frac{1}{8}\sum_i k_i^3 - \frac{1}{8} \sum_{i\neq j} k_i k_j^2 +
\frac{1}{8}k_1 k_2 k_3  + \frac{1}{2 K} \sum_{i<j} k_i^2 k_j^2-
\frac{1}{2 K^2} \sum_{i\neq j}k_i^2 k_j^3\right] \right\},
\end{align}
where we have used expressions (\ref{intcsinf1}-\ref{intcsinf3})
and (\ref{deffnl}). Note that except for the terms containing an
explicit dependence on the flow parameters, $\gamma = \epsilon + s
- 1$, equation (\ref{defgam}), expression (\ref{ngamplitude}) is
identical to the one obtained in equation (4.15) of
\cite{Noller:2011hd}; as pointed out in the Introduction, such
contributions are absent in both DBI and nonminimal bimetric
models, for they come from the $\dot\zeta^3$ contribution in
action (\ref{3ordaction}). In the next two sections, I discuss the
modifications induced by such contributions with regard to the
full superluminal amplitude found in \cite{Noller:2011hd}.

\subsection{Size of non-Gaussian amplitudes, $f_{\rm NL}$}
\label{sec:sizengamplitudes}

The parameter that characterizes the three-point amplitude,
$f_{\rm NL}$, is defined as \cite{Komatsu:2001rj}
\begin{eqnarray}
\zeta = \zeta_g(x) + \frac{3}{5}f_{\rm NL}\zeta_g^2;
\end{eqnarray}
following \cite{Khoury:2008wj}, we define $f_{\rm NL}$ at
$k_1=k_2=k_3=K/3$, so that
\begin{eqnarray}
\label{deffnl}
f_{\rm NL} = 30\frac{{\cal A}_{k_1=k_2=k_3}}{K^3}.
\end{eqnarray}

For the cuscuton-like models, substituting (\ref{ngamplitude})
into (\ref{deffnl}), we get
\begin{align}
\label{fnltach}
f^{c}_{\rm NL} = \left(\frac{1}{54}\right)^{n_s -1}
\left\{\frac{5}{81}\left(\epsilon + s - 1\right)+\frac{35}{108} +
(n_s -1)\left[
\frac{5}{54}\left(\epsilon + s - 1\right)-\frac{25}{27}\right]
\right\},
\end{align}
where superscript $c$ stands for {\it{cuscuton}}. As for
superluminal disformal bimetric theories, the corresponding size
of non-Gaussianities $f_{\rm NL}$ is given by equation (3.16) in
\cite{Noller:2011hd},
\begin{eqnarray}
f_{\rm NL} \sim 0.28-0.04\epsilon-(1.19+0.08\epsilon)(n_s-1);
\end{eqnarray}
then, for a red-tilted spectrum $n_s\sim 0.96$, we find, for
$\epsilon = 0.01$ and $\epsilon = 0.3$ (these are the $\epsilon$
values they have chosen to plot Figure 6),
\begin{eqnarray}
\label{bimsig} f_{\rm NL} \sim 0.327,\, \, \, f_{\rm NL} \sim
0.317,
\end{eqnarray}
respectively; in the case of matter-dominated ($\epsilon = 3/2$)
and radiation-dominated ($\epsilon = 2$) cuscuton-like
tachyacoustic models one gets
\begin{eqnarray}
\label{cuscsig}
f^{c}_{\rm NL} \sim 0.266,~~~~~f^{c}_{\rm NL} \sim
0.232
\end{eqnarray}
respectively. Then, cuscuton-like models also produce
non-Gaussianities of the same order of magnitude as of the other
superluminal models, albeit with different magnitudes, as can be
seen from the results (\ref{bimsig}) and (\ref{cuscsig}). Such
differences come from the fact that the $\dot\zeta^3$
contribution, present in cuscuton-like models, is not negligible
and possesses negative sign (for its $\gamma$ coefficient, given
by equation (\ref{defgam}), is {\it negative} for the flow
parameters $\epsilon$ and $s$ chosen), so that it {\it reduces}
not only the size of the non-Gaussianities, but their shapes as
well.

\subsection{Shapes of non-Gaussian amplitudes}
\label{sec:shape}

Following the literature \cite{Babich:2004gb}, I discuss next the
shape of the amplitude at fixed $K$, by means of the dimensionless
ratio ${\cal{A}}_{\bar{c}_s \to \infty}/k_1k_2k_3$. It is
convenient to change the variables $k_2$ and $k_3$ into the
dimensionless ones $x_2=k_2/k_1$ and $x_3=k_3/k_1$; then, the
amplitude to be studied is ${\cal{A}}_{\bar{c}_s \to
\infty}\left(1,x_2,x_3\right)/x_2x_3$ in the $x_2-x_3$ plane.
Without any loss of generality the momenta are ordered such that
$x_3<x_2<1$; the triangle inequality implies $x_2+x_3>1$. To avoid
plotting the same configuration twice, the amplitude is set to
zero outside the range $1-x_2 < x_3 x_2$.

\begin{figure}[!htb]
\begin{minipage}[b]{0.40\linewidth}
\includegraphics[width=\linewidth]{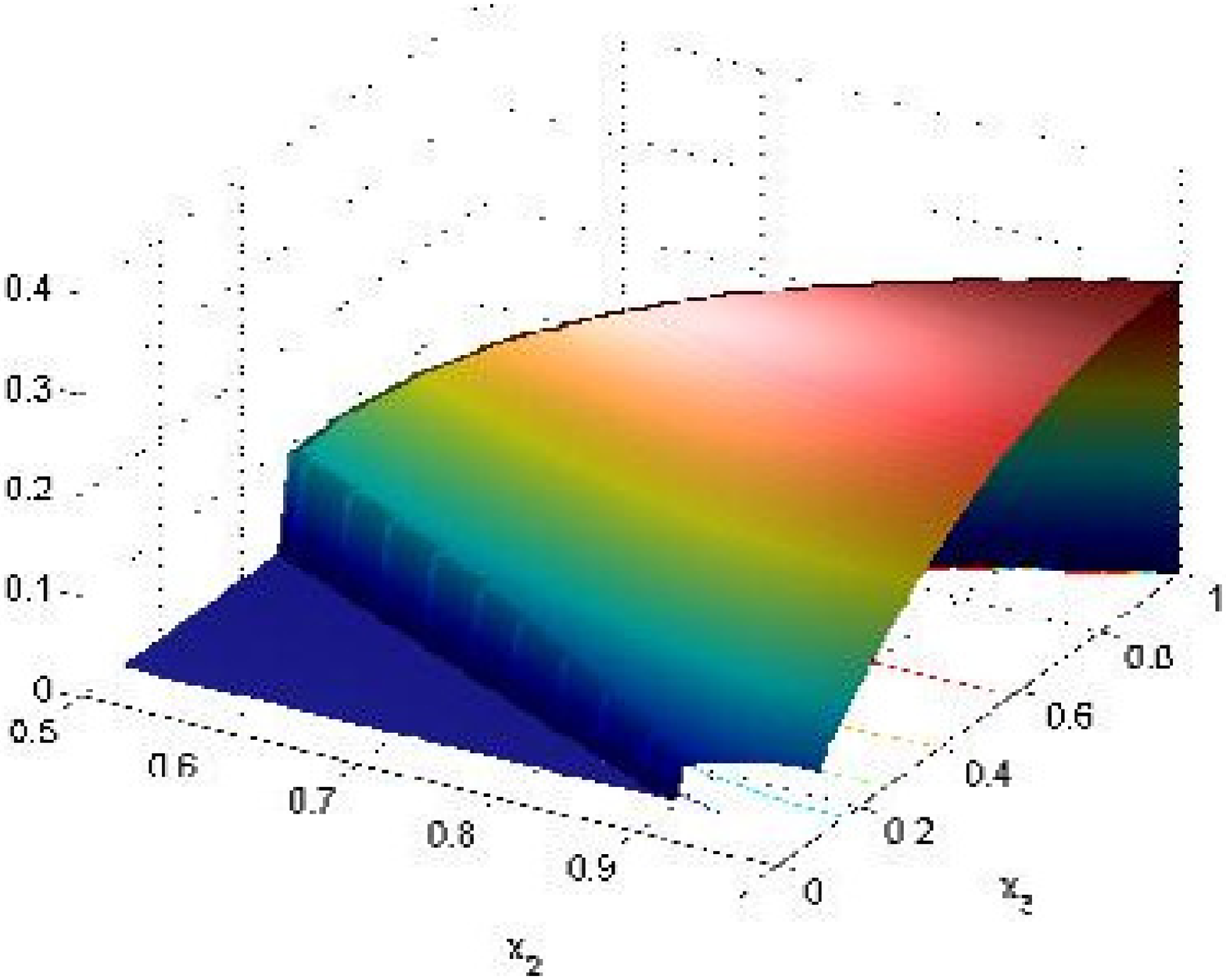}
\caption{Non-Gaussian amplitude ${\cal{A}}_{\bar{c}_s \to
\infty}\left(1,x_2,x_3\right)/x_2x_3$ for $n_s=1$ without the
$\dot\zeta^3$ contribution.\label{fig:sci}}
\end{minipage} \hfill
\begin{minipage}[b]{0.40\linewidth}
\includegraphics[width=\linewidth]{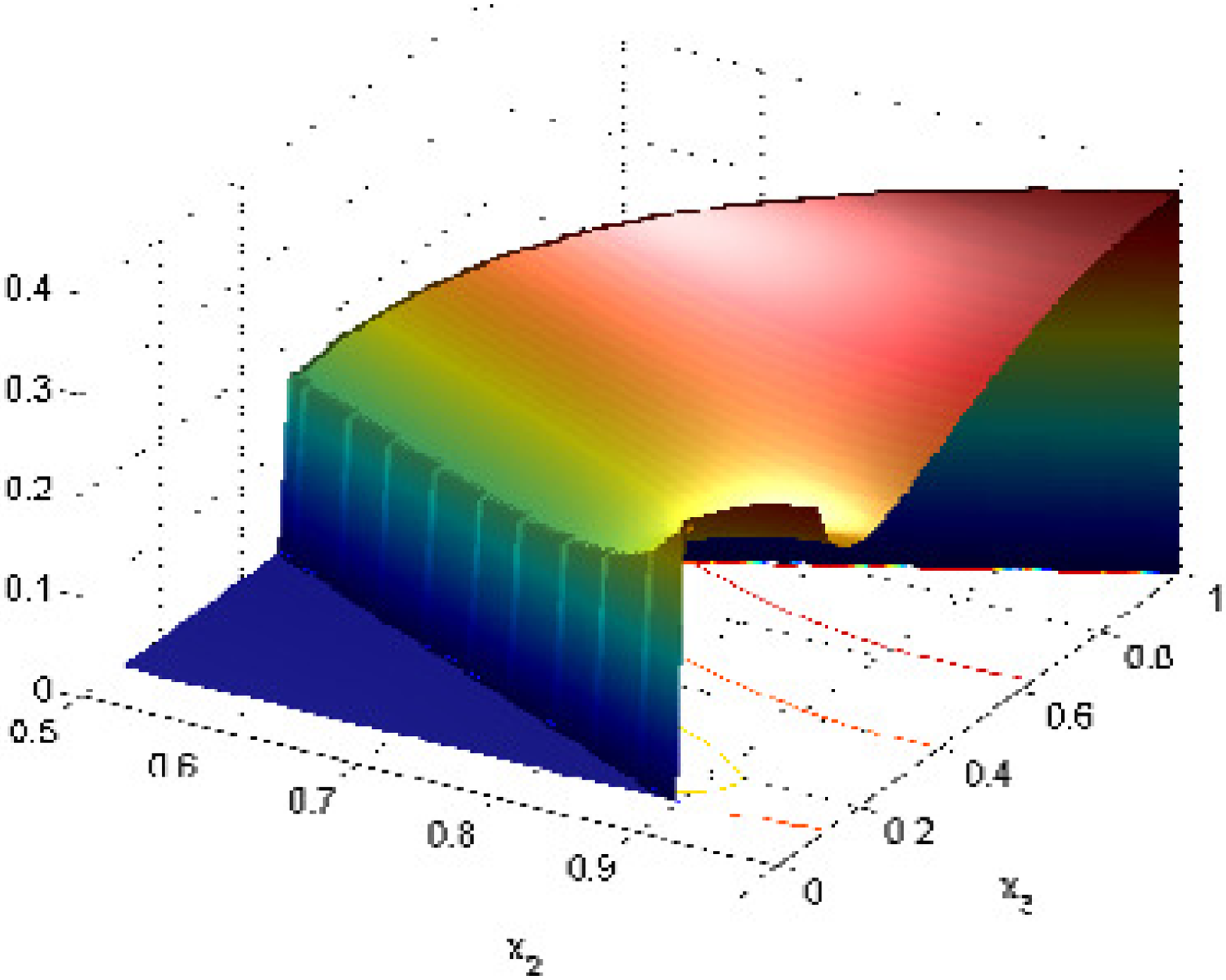}
\caption{Non-Gaussian amplitude for the same model, but with
$n_s=0.96$.\label{fig:bimnsci}}
\end{minipage}
\end{figure}

For the sake of comparison, it is instructive to discuss briefly
the results obtained in \cite{Magueijo:2010zc}. First, the full
amplitude generated is given by their equation (4.15),
\begin{align}
\label{ngamplbim} {\cal A}_{\bar{c}_s \to \infty} \ = &\
\left(\frac{k_1 k_2 k_3}{2 K^3}\right)^{n_s -1}
\left\{
- \frac{1}{8}\sum_i k_i^3 + \frac{1}{K} \sum_{i<j} k_i^2 k_j^2 -
\frac{1}{2 K^2} \sum_{i\neq j}k_i^2 k_j^3\, \right. \nonumber \\
&\ + (n_s -1)\left.\left[
- \frac{1}{8}\sum_i k_i^3 - \frac{1}{8} \sum_{i\neq j} k_i k_j^2 +
\frac{1}{8}k_1 k_2 k_3  + \frac{1}{2 K} \sum_{i<j} k_i^2 k_j^2-
\frac{1}{2 K^2} \sum_{i\neq j}k_i^2 k_j^3\right] \right\},
\end{align}
whose shape is depicted in Figure \ref{fig:sci}, for the
scale-invariant limit, and in Figure \ref{fig:bimnsci} for a
red-tilted spectrum.

%\begin{center}
%\begin{figure}[!htb]
%\centering
%\includegraphics[scale=0.5]{scale_invariant.eps}
%\includegraphics[scale=0.5]{bimetric_near_sc_inv.eps}
%\caption{Non-Gaussian shape . \label{fig:shapbim}}
%\end{figure}
%\end{center}

\begin{center}
\begin{figure}[!htb]
\centering
\includegraphics[scale=0.30]{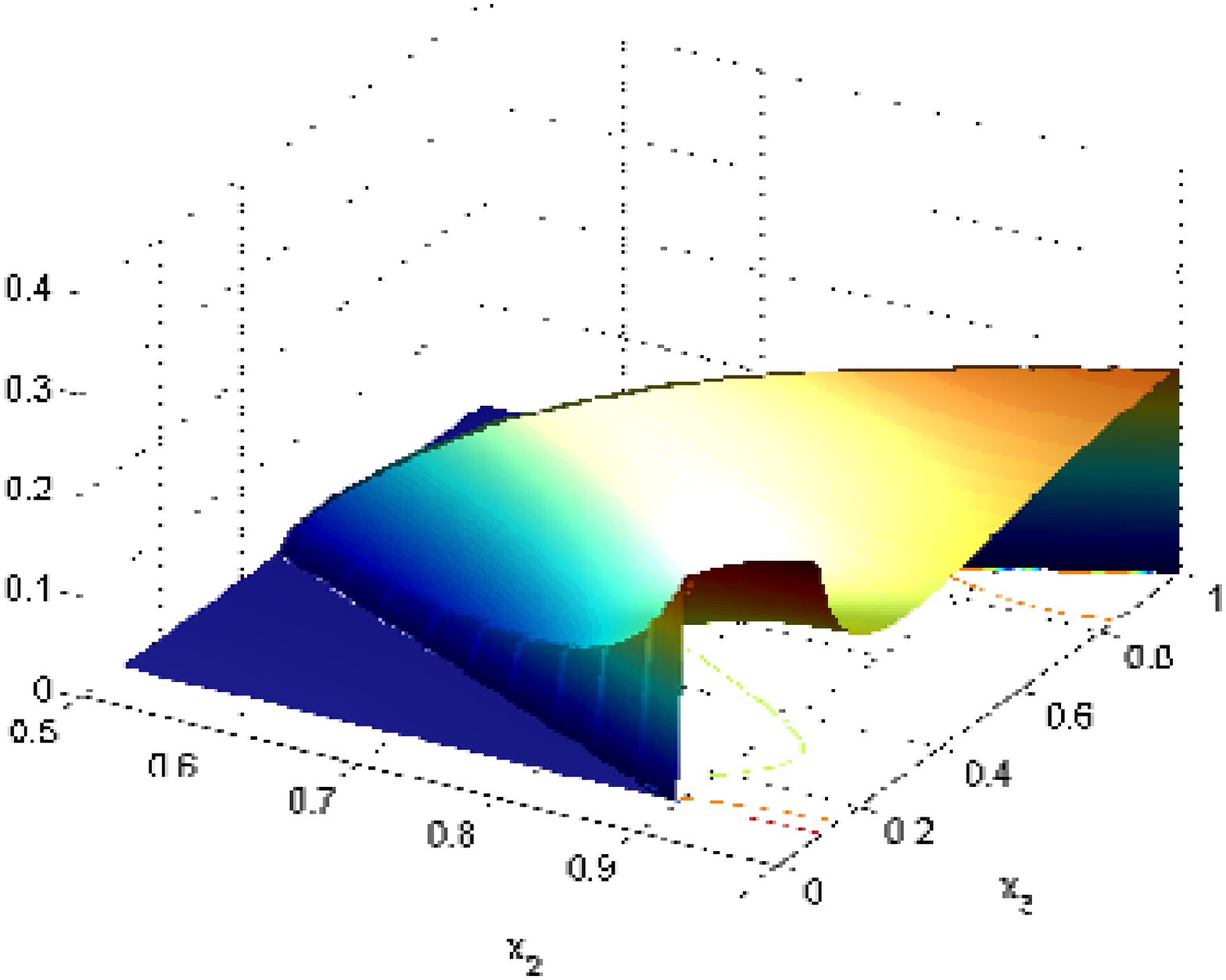}
\includegraphics[scale=0.30]{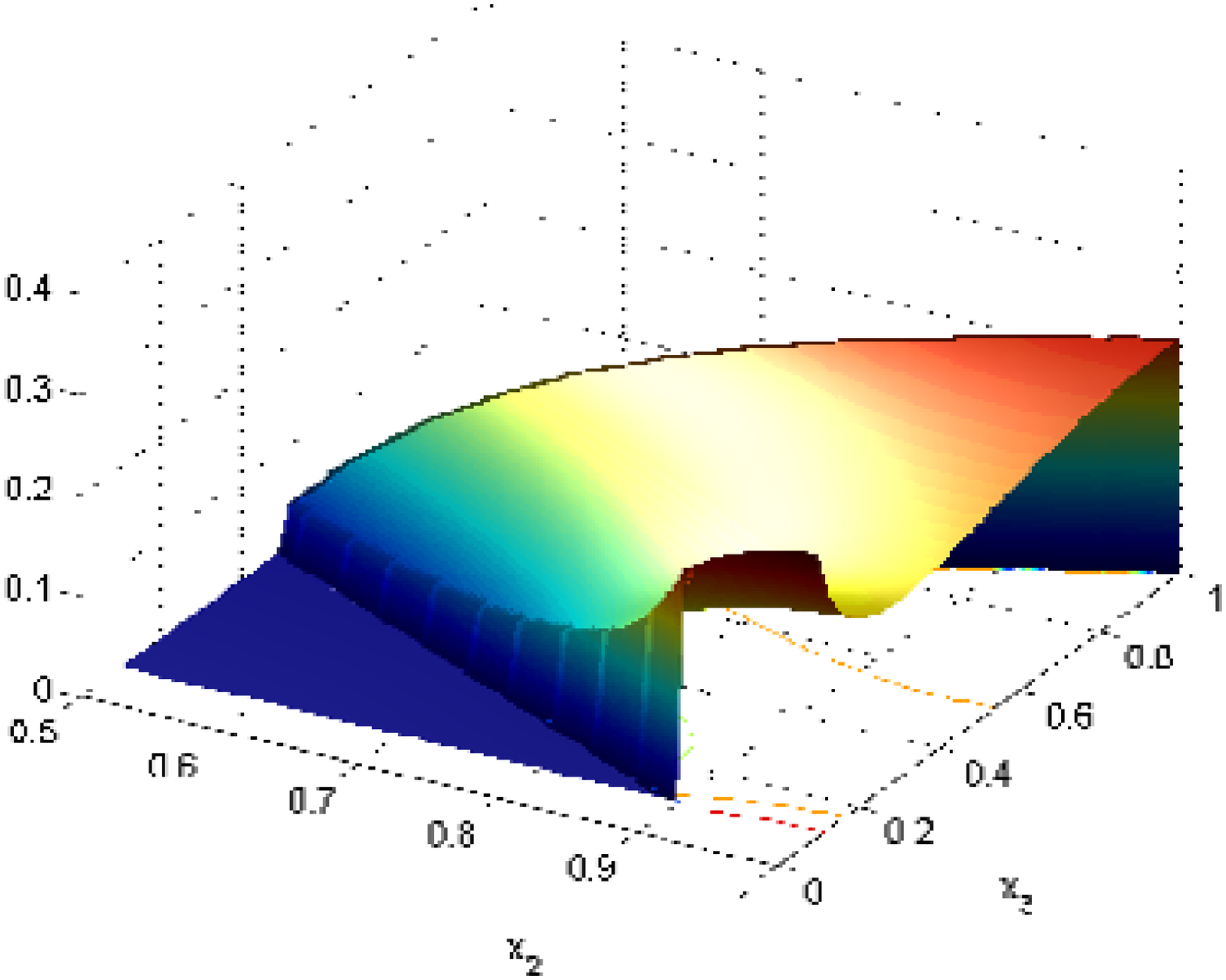}
\caption{Non-Gaussian amplitude ${\cal{A}}_{\bar{c}_s \to
\infty}\left(1,x_2,x_3\right)/x_2x_3$ for the cuscuton-like model
with $n_s=0.96$ with a radiative equation of state (left) and with
a matter equation of state (right).\label{fig:cuscradmat}}
\end{figure}
\end{center}

As for the cuscuton-like models, we see in Figure
\ref{fig:cuscradmat} that the extra contribution coming from the
$\dot\zeta^3$ term modifies the shape dependence. In bimetric
models, for instance, the amplitude peaks in the equilateral limit
$k_1 = k_2 = k_3$ for both scale-invariant and red-tilted spectra;
in the latter, besides, the amplitude also peaks in the local
limit $k_3\ll k_1 = k_2$, which dominates over the equilateral
mode. Cuscuton-like models share the same features, as can be seen
in the Figure \ref{fig:cuscradmat}; also, comparing Figures
\ref{fig:bimnsci} and \ref{fig:cuscradmat} it is clear that the
$\dot\zeta^3$-terms contributes to reduce the magnitude of the
non-Gaussian amplitudes, specially in the equilateral
configuration.

This effect is stronger in the radiative cuscuton-like model,
which is expected, since the combination (\ref{defgam}) yields
$\gamma\sim -2.9$ for a radiative equation of state with $\epsilon
=2$ and $s\sim -3.9$, whereas $\gamma\sim -2.4$ for a matter
equation of state with $\epsilon =3/2$ and $s\sim -2.9$. Hence,
bimetric and cuscuton-like models, despite their similarities, can
be clearly distinguished by means of the different results
obtained for their sizes and shapes of non-Gaussianity.

\section{Conclusions}
\label{sec:concl}

Among the models proposed to solve the puzzles of the very-early
universe, those with noncanonical Lagrangians possess an
interesting property, the possibility of a varying speed of sound.
For subluminal models, as in DBI inflation, a small speed of sound
can produce large non-Gaussian amplitudes as $f_{\rm NL}\sim
-100$, for example; in the superluminal case, for instance, as in
the nonminimal bimetric model, large non-Gaussianities are
suppressed, and the predicted size of non-Gaussianities amounts to
be around $f_{\rm NL}\sim {\cal O}(1)$. Such model admits a
nearly-scale invariant spectrum, which induces distortions to the
equilateral shape of non-Gaussianities, and hence making it
distinguishable from other noncanonical subluminal models.

The cuscuton-like models proposed in \cite{Bessada:2009ns} also
possess a superluminal speed of sound, with the advantage of being
also noninflationary. This means that, unlike inflationary models,
we can choose values greater than one for the flow parameter
$\epsilon$, thus eliminating the reheating period for a scalar
field with a radiative equation of state. Since they lead to a
nearly-scale invariant spectrum, it is natural to take one step
further and investigate their non-Gaussian signatures, which was
the main subject of this work. After deriving the amplitudes for
each term in the third-order action, equation (\ref{3ordaction}),
I found that the $\dot\zeta^3$ contribution in cuscuton-like
models is nonzero, unlike in the nonminimal bimetric model, due to
its DBI-type action. To check whether this contribution induces
any significant deviation from the amplitude of bimetric models, I
derived an expression for the size of the non-Gaussian amplitudes,
given by equation (\ref{fnltach}), and found that $f_{\rm NL}\sim
{\cal O}(1)$, which is of the same order as bimetric models.
However, a small difference in the corresponding values of the
$f_{\rm NL}$ parameter was found (equations (\ref{bimsig}) and
(\ref{cuscsig})), due exactly to the negative coefficient $\gamma$
emerging from the $\dot\zeta^3$ contribution to the three-point
amplitude in the cuscuton-like models. Such extra terms reduce the
size of non-Gaussianities, as well as the height of their shapes
in comparison to bimetric models. Also, the equilateral
configuration becomes subdominant, whereas the ``squeezed"
triangle mode becomes relevant.

Then, cuscuton-like models share similar features with
superluminal bimetric models, but they have distinct non-Gaussian
signatures, which is evident from the comparison of the shapes of
their corresponding amplitudes. Due to the low values predicted
for $f_{\rm NL}$, which are still well below the upcoming
constraints to be posed by Planck satellite measurements, it is
not yet possible to falsify such models observationally, but only
those predicting huge non-Gaussian amplitude sizes.

\begin{acknowledgments}
I thank Brazilian agency FAPESP, grant 2009/15612-6 for financial
support at the beginning of this project. I also thank Will
Kinney, Oswaldo Miranda and Jos\'e Ademir Salles de Lima for
helpful suggestions.

\end{acknowledgments}


\begin{thebibliography}{99}

\bibitem{inflation}
   A.~H.~Guth,
   %``The Inflationary Universe: A Possible Solution To The Horizon And Flatness Problems,''
   Phys.\ Rev.\ D {\bf 23}, 347 (1981).
%%CITATION = PHRVA,D23,347;%%
   A.~D.~Linde,
   %``A New Inflationary Universe Scenario: A Possible Solution Of The Horizon, Flatness,
   %Homogeneity, Isotropy And Primordial Monopole Problems,''
   Phys.\ Lett.\ B {\bf 108}, 389 (1982).
   %%CITATION = PHLTA,B108,389;%%
   A.~Albrecht and P.~J.~Steinhardt,
   %``Cosmology For Grand Unified Theories With Radiatively Induced Symmetry Breaking,''
   Phys.\ Rev.\ Lett.\  {\bf 48}, 1220 (1982).
   %%CITATION = PRLTA,48,1220;%%

\bibitem{prebigbang}
G.~Veneziano,
  %``Scale Factor Duality For Classical And Quantum Strings,''
  Phys.\ Lett.\  B {\bf 265}, 287 (1991);
  M.~Gasperini and G.~Veneziano,
  %``Pre - big bang in string cosmology,''
  Astropart.\ Phys.\  {\bf 1}, 317 (1993)
  [arXiv:hep-th/9211021];
  M.~Gasperini and G.~Veneziano,
  %``The pre-big bang scenario in string cosmology,''
  Phys.\ Rept.\  {\bf 373}, 1 (2003)
  [arXiv:hep-th/0207130].
  V.~Bozza, M.~Gasperini, M.~Giovannini and G.~Veneziano,
  %``Assisting pre-big bang phenomenology through short-lived axions,''
  Phys.\ Lett.\  B {\bf 543}, 14 (2002)
  [arXiv:hep-ph/0206131];
  %%CITATION = PHLTA,B543,14;%%
V.~Bozza and G.~Veneziano,
  %``Regular two-component bouncing cosmologies and perturbations therein,''
  JCAP {\bf 0509}, 007 (2005)
  [arXiv:gr-qc/0506040].
  %%CITATION = JCAPA,0509,007;%%

\bibitem{ekpyrotic}
J.~Khoury, B.~A.~Ovrut, P.~J.~Steinhardt and N.~Turok,
  %``The ekpyrotic universe: Colliding branes and the origin of the hot big
  %bang,''
  Phys.\ Rev.\  D {\bf 64}, 123522 (2001)
  [arXiv:hep-th/0103239];
  %%CITATION = PHRVA,D64,123522;%%
P. Steinhardt and N. Turok, Science 296: 1436-1439, 2002.
J.~L.~Lehners, P.~McFadden, N.~Turok and P.~J.~Steinhardt,
  %``Generating ekpyrotic curvature perturbations before the big bang,''
  Phys.\ Rev.\  D {\bf 76}, 103501 (2007)
  [arXiv:hep-th/0702153];
  E.~I.~Buchbinder, J.~Khoury and B.~A.~Ovrut,
  %``New Ekpyrotic Cosmology,''
Phys.\ Rev.\  D {\bf 76}, 123503(2007) ÊÊÊ [arXiv:hep-th/0702154];
P.~Creminelli and L.~Senatore,
  %``A smooth bouncing cosmology with scale invariant spectrum,''
  JCAP {\bf 0711}, 010 (2007)
  [arXiv:hep-th/0702165];
  %%CITATION = JCAPA,0711,010;%%
  K.~Koyama and D.~Wands,
  %``Ekpyrotic collapse with multiple fields,''
  JCAP {\bf 0704}, 008 (2007)
  [arXiv:hep-th/0703040];
  %%CITATION = JCAPA,0704,008;%%
K.~Koyama, S.~Mizuno and D.~Wands,
  %``Curvature perturbations from ekpyrotic collapse with multiple fields,''
  Class.\ Quant.\ Grav.\  {\bf 24}, 3919 (2007)
  [arXiv:0704.1152 [hep-th]].
  %%CITATION = CQGRD,24,3919;%%

\bibitem{quantumcosmology}
J.~Acacio de Barros, N.~Pinto-Neto, and M.~A.~Sagioro-Leal, Phys.
Lett. A {\bf 241}, 229 (1998). R.~Colistete Jr., J.~C.~Fabris, and
N.~Pinto-Neto, \prd {\bf 62}, 083507 (2000). F.G. Alvarenga, J.C.
Fabris, N.A. Lemos and G.A. Monerat, Gen.Rel.Grav. {\bf 34}, 651
(2002).

\bibitem{noncanonical}
%\bibitem{ArmendarizPicon:1999rj}
  C.~Armendariz-Picon, T.~Damour and V.~F.~Mukhanov,
  %``k-Inflation,''
  Phys.\ Lett.\  B {\bf 458}, 209 (1999)
  [arXiv:hep-th/9904075].
  %%CITATION = PHLTA,B458,209;%%
%\bibitem{Silverstein:2003hf}
  E.~Silverstein and D.~Tong,
  %``Scalar Speed Limits and Cosmology: Acceleration from D-cceleration,''
  Phys.\ Rev.\  D {\bf 70}, 103505 (2004)
  [arXiv:hep-th/0310221].
  %%CITATION = PHRVA,D70,103505;%%
%\bibitem{ArmendarizPicon:2006if}
   C.~Armendariz-Picon,
   %``Near scale invariance with modified dispersion relations,''
   JCAP {\bf 0610}, 010 (2006)
   [arXiv:astro-ph/0606168].
   %%CITATION = JCAPA,0610,010;%%
%\bibitem{Piao:2006ja}
  Y.~S.~Piao,
  %``Seeding of Primordial Perturbations During a Decelerated Expansion,''
  Phys.\ Rev.\  D {\bf 75}, 063517 (2007)
  [arXiv:gr-qc/0609071].
  %%CITATION = PHRVA,D75,063517;%%
%\bibitem{Magueijo:2008pm}
  J.~Magueijo,
  %``Speedy sound and cosmic structure,''
  Phys.\ Rev.\ Lett.\  {\bf 100}, 231302 (2008)
  [arXiv:0803.0859 [astro-ph]].
  %%CITATION = PRLTA,100,231302;%%
%\bibitem{Magueijo:2008sx}
  J.~Magueijo,
  %``Bimetric varying speed of light theories and primordial fluctuations,''
  Phys.\ Rev.\  D {\bf 79}, 043525 (2009)
  [arXiv:0807.1689 [gr-qc]].
  %%CITATION = PHRVA,D79,043525;%%
%\bibitem{Piao:2008ip}
  Y.~S.~Piao,
  %``On Primordial Density Perturbation and Decaying Speed of Sound,''
  arXiv:0807.3226 [gr-qc].
  %%CITATION = ARXIV:0807.3226;%%

\bibitem{Bessada:2009ns}
D.~Bessada, W.~H.~Kinney, D.~Stojkovic and J.~Wang,
  %``Tachyacoustic Cosmology: An Alternative to Inflation,''
  Phys.\ Rev.\ D {\bf 81}, 043510 (2010)
  [arXiv:0908.3898 [astro-ph.CO]].
  %%CITATION = ARXIV:0908.3898;%%

\bibitem{Bessada:2009pe}
  D.~Bessada, W.~H.~Kinney and K.~Tzirakis,
  %``Inflationary potentials in DBI models,''
  JCAP {\bf 0909}, 031 (2009)
  [arXiv:0907.1311 [gr-qc]].
  %%CITATION = ARXIV:0907.1311;%%

\bibitem{Alishahiha:2004eh}
  M.~Alishahiha, E.~Silverstein and D.~Tong,
  %``DBI in the sky,''
  Phys.\ Rev.\ D {\bf 70}, 123505 (2004)
  [hep-th/0404084].
  %%CITATION = HEP-TH/0404084;%%

\bibitem{Seery:2005wm}
  D.~Seery and J.~E.~Lidsey,
  %``Primordial non-Gaussianities in single field inflation,''
  JCAP {\bf 0506}, 003 (2005)
  [astro-ph/0503692].
  %%CITATION = ASTRO-PH/0503692;%%

\bibitem{Chen:2006nt}
  X.~Chen, M.~-x.~Huang, S.~Kachru and G.~Shiu,
  %``Observational signatures and non-Gaussianities of general single field inflation,''
  JCAP {\bf 0701}, 002 (2007)
  [hep-th/0605045].
  %%CITATION = HEP-TH/0605045;%%

\bibitem{Khoury:2008wj}
  J.~Khoury and F.~Piazza,
  %``Rapidly-Varying Speed of Sound, Scale Invariance and Non-Gaussian Signatures,''
  JCAP {\bf 0907}, 026 (2009)
  [arXiv:0811.3633 [hep-th]].
  %%CITATION = ARXIV:0811.3633;%%

\bibitem{Magueijo:2010zc}
  J.~Magueijo, J.~Noller and F.~Piazza,
  %``Bimetric structure formation: non-Gaussian predictions,''
  Phys.\ Rev.\ D {\bf 82}, 043521 (2010)
  [arXiv:1006.3216 [astro-ph.CO]].
  %%CITATION = ARXIV:1006.3216;%%

  \bibitem{Noller:2011hd}
  J.~Noller and J.~Magueijo,
  %``Non-Gaussianity in single field models without slow-roll,''
  Phys.\ Rev.\ D {\bf 83}, 103511 (2011)
  [arXiv:1102.0275 [astro-ph.CO]].
  %%CITATION = ARXIV:1102.0275;%%

\bibitem{Komatsu:2001rj}
  E.~Komatsu and D.~N.~Spergel,
  %``Acoustic signatures in the primary microwave background bispectrum,''
  Phys.\ Rev.\ D {\bf 63}, 063002 (2001)
  [astro-ph/0005036].
  %%CITATION = ASTRO-PH/0005036;%%

\bibitem{Maldacena:2002vr}
  J.~M.~Maldacena,
  %``Non-Gaussian features of primordial fluctuations in single field inflationary models,''
  JHEP {\bf 0305}, 013 (2003)
  [astro-ph/0210603].
  %%CITATION = ASTRO-PH/0210603;%%

\bibitem{Bessada:2012zx}
  D.~Bessada and W.~H.~Kinney,
  ``Attractor Solutions in Tachyacoustic Cosmology,'', To appear
  in Physical Review D
  arXiv:1206.2711 [gr-qc].
  %%CITATION = ARXIV:1206.2711;%%

\bibitem{Afshordi:2006ad}
  N.~Afshordi, D.~J.~H.~Chung and G.~Geshnizjani,
  %``Cuscuton: A Causal Field Theory with an Infinite Speed of Sound,''
  Phys.\ Rev.\ D {\bf 75}, 083513 (2007)
  [hep-th/0609150].
  %%CITATION = HEP-TH/0609150;%%

\bibitem{vsl}
  J.~W.~Moffat,
  %``Superluminary universe: A Possible solution to the initial value problem in cosmology,''
  Int.\ J.\ Mod.\ Phys.\ D {\bf 2}, 351 (1993)
  [gr-qc/9211020].
  %%CITATION = GR-QC/9211020;%%

%\bibitem{Albrecht:1998ir}
  A.~Albrecht and J.~Magueijo,
  %``A Time varying speed of light as a solution to cosmological puzzles,''
  Phys.\ Rev.\ D {\bf 59}, 043516 (1999)
  [astro-ph/9811018].
  %%CITATION = ASTRO-PH/9811018;%%

%\bibitem{Magueijo:2003gj}
  J.~Magueijo,
  %``New varying speed of light theories,''
  Rept.\ Prog.\ Phys.\  {\bf 66}, 2025 (2003)
  [astro-ph/0305457].
  %%CITATION = ASTRO-PH/0305457;%%

\bibitem{Bruneton:2006gf}
  J.~-P.~Bruneton,
  %``On causality and superluminal behavior in classical field theories: Applications to k-essence theories and MOND-like theories of gravity,''
  Phys.\ Rev.\ D {\bf 75}, 085013 (2007)
  [gr-qc/0607055].
  %%CITATION = GR-QC/0607055;%%

\bibitem{Babichev:2007dw}
  E.~Babichev, V.~Mukhanov and A.~Vikman,
  %``k-Essence, superluminal propagation, causality and emergent geometry,''
  JHEP {\bf 0802}, 101 (2008)
  [arXiv:0708.0561 [hep-th]].
  %%CITATION = ARXIV:0708.0561;%%

\bibitem{Bean:2008ga}
  R.~Bean, D.~J.~H.~Chung and G.~Geshnizjani,
  %``Reconstructing a general inflationary action,''
  Phys.\ Rev.\  D {\bf 78}, 023517 (2008)
  [arXiv:0801.0742 [astro-ph]].
  %%CITATION = PHRVA,D78,023517;%%

\bibitem{Garriga:1999vw}
  J.~Garriga and V.~F.~Mukhanov,
  %``Perturbations in k-inflation,''
  Phys.\ Lett.\  B {\bf 458}, 219 (1999)
  [arXiv:hep-th/9904176].
  %%CITATION = PHLTA,B458,219;%%

\bibitem{Babich:2004gb}
  D.~Babich, P.~Creminelli and M.~Zaldarriaga,
  %``The Shape of non-Gaussianities,''
  JCAP {\bf 0408}, 009 (2004)
  [astro-ph/0405356].
  %%CITATION = ASTRO-PH/0405356;%%


\end{thebibliography}
\end{document}